# Modeling European Options: A simulation study

Aishwarya B U (1PE10CS006)
Mohammed Saaqib A (1PE10CS053)
Rajashree H R (1PE10CS074)
Vigasini B (1PE10CS111)

# 1. Title:

Modeling European Options: A simulation study

# 2. Abstract:

Option contracts can be valued by using the Black-Scholes equation, a partial differential equation with initial conditions. An exact solution for European style options is known. The computation time and the error need to be minimized simultaneously. In this paper, the authors have solved the Black-Scholes equation by employing a reasonably accurate implicit method. Options with known analytic solutions have been evaluated. Furthermore, an overall second order accurate space and time discretization has been accomplished in this paper.

**2.1 Keywords:** Computational finance, implicit methods, finite differences, call/put options.

# 3. Background Study and Survey:

**3.1 The Black-Scholes Equation:**

**3.1.1 Assumptions**:

The fundamental assumption made about the random movement of asset prices for a more flexible hypothesis.
► We assume that present price is a full reflection of the past history experienced. And does not contain any more information;
► We assume that we would obtain immediate response for updated information about an asset.

**3.1.2 Model:**

Asset price modeling identifies modeling the arrival of new information which affects the price of an asset. Absolute change in an asset is a relative measurement of the change in price. This is a better indicator of its size aspect than absolute measures.

Let's consider that at time $t$, the asset price is $S$. Let us consider a small subsequent time interval $dt$, during which $S$ changes to $S + dS$. Now, there is deterministic and anticipated return represented by $\mu dt$, where $\mu$ is known as drift. The other factor affecting the asset price is the random change in the asset price in response to external effects. It is represented by a random sample drawn from a normal distribution with mean zero denoted as $\sigma dX$,

Putting the above two factors together we obtain a stochastic differential equation,

$$\frac{dS}{S} = \sigma dX + \mu dt,  \qquad (1)$$

which is the mathematical representation for generating asset prices[1,4].

The equation (1) gives interesting and important information concerning the behavior of $S$ in a probabilistic sense. At time $t = t'$, suppose the price is $S'$, and then $S'$ will be distributed about $S_0$ with a probability density function. The future asset price $S'$ is thus most likely to be close to $S_0$ and less likely to be far away. Thus, the equation generates time series – each time the series is restarted a different path results. Each path is called a **realization** of the random walk.

The **Black-Scholes Partial Differential Equation,** the derivation of which is well-known, is given below

$$\frac{\partial V}{\partial t} + \frac{1}{2}\sigma^2 S^2 \frac{\partial^2 V}{\partial S^2} + rS\frac{\partial V}{\partial S} - rV = 0  \qquad (2)$$

If $S > E$ at expiry, then in financial sense, for call option, there will be profit $S - E$, handing over an amount $E$, to obtain an asset worth, while, if $S < E$, then there will be loss of $E - S$. Thus the value of the call option at expiry can be written as

$$C(S,T) = \max(S - E, 0)  \qquad (3)$$

As the time tends to expiry date the value of call option approaches (6), it is known as **pay-off function** for European Call Option. This is known as the final condition of PDE (12).

Now, from (1), if $S = 0$ then $dS = 0$ which means pay-off is also zero. Thus, the call option is worthless on $S = 0$ even if there is long time to expiry. Hence, we have

$$C(0,t) = 0  \qquad (4)$$

And if $S \to \infty$ i.e. the asset price increases without bound it becomes ever more likely that the option will be exercised and the magnitude of the exercise price becomes less and less important. Thus as $S \to \infty$ the value of the option becomes that of the asset and so

$$C(S,t) \sim S \text{ as } S \to \infty  \qquad (5)$$

Thus, the Black Scholes equation and boundary condition for a **European Call Option** is given by equations (2)-(5).

For European Put Option, the final condition is the payoff
$$P(S,T) = \max(E - S, 0) \tag{6}$$

And the initial condition $P(0,t)$ is determined by calculating the present value of an amount $E$ received at time $T$. For time-dependent interest rate, the boundary condition at $S = 0$ is

$$P(0,t) = Ee^{-\int_t^T r(T)dT} \tag{7}$$

And,
$$P(S,T) \to 0 \text{ as } S \to \infty \tag{8}$$

Thus, the Black Scholes equation and boundary conditions for a **European Put Option** are given by equations (2), (6)-(8).

The analytical solution to the BSDE is not the focal point of the paper but given below, nonetheless as [2]
$$C(S,t) = SN(d_1) - Ee^{-r(T-t)}N(d_2) \tag{9}$$

where
$$d_1 = \frac{\log(S/E) + \left(r + \frac{1}{2}\sigma^2\right)(T-t)}{\sigma(\sqrt{T-t})}$$

$$d_2 = \frac{\log(S/E) + \left(r - \frac{1}{2}\sigma^2\right)(T-t)}{\sigma(\sqrt{T-t})}$$

The corresponding calculation for a European Call Option follows similar lines.

### 3.2 Analytical Solution:

$$\frac{\partial V}{\partial t} + \frac{1}{2}\sigma^2 S^2 \frac{\partial^2 V}{\partial S^2} + rS\frac{\partial V}{\partial S} - rV = 0$$

with initial condition
$$C(S,T) = \max(S - E, 0)$$
$$C(0,t) = 0$$

and boundary condition

$$C(S,t) \sim S \text{ as } S \to \infty$$

For *European Call Option,* suppose

$$S = Ee^x, \quad t = T - \tau \Big/ \frac{1}{2}\sigma^2, \quad C = Ev(x,\tau) \tag{i}$$

$$V \to v \ \& \ S \to x$$

Then the **Black-Scholes Partial Differential Equation** results in

$$\frac{\partial v}{\partial \tau} = \frac{\partial^2 v}{\partial x^2} + (k-1)\frac{\partial v}{\partial x} - kv \tag{ii}$$

Where $k = r \Big/ \frac{1}{2}\sigma^2$. So, the initial condition changes to

$$v(x,0) = \max(e^x - 1, 0)$$

Using the method of change of variable, we construct

$$v = e^{\alpha x + \beta \tau} u(x,\tau),$$

for some constants $\alpha$ and $\beta$, subsequent differentiation yields

$$\frac{\partial v}{\partial \tau} = e^{\alpha x + \beta \tau} \frac{\partial u}{\partial \tau} + \beta e^{\alpha x + \beta \tau} u(x,\tau)$$

$$\frac{\partial v}{\partial \tau} = e^{\alpha x + \beta \tau} \frac{\partial u}{\partial \tau} + \alpha e^{\alpha x + \beta \tau} u(x,\tau)$$

$$\frac{\partial^2 v}{\partial x^2} = e^{\alpha x + \beta \tau} \frac{\partial^2 u}{\partial x^2} + \alpha e^{\alpha x + \beta \tau} \frac{\partial u}{\partial x} + \alpha^2 e^{\alpha x + \beta \tau} u + \alpha e^{\alpha x + \beta \tau} \frac{\partial u}{\partial x}$$

$$\frac{\partial^2 v}{\partial x^2} = \alpha^2 e^{\alpha x + \beta \tau} u + 2\alpha e^{\alpha x + \beta \tau} \frac{\partial u}{\partial x} + e^{\alpha x + \beta \tau} \frac{\partial^2 u}{\partial x^2}$$

$$e^{\alpha x + \beta \tau}\left(\beta u + \frac{\partial u}{\partial \tau}\right) = e^{\alpha x + \beta \tau}\left(\alpha^2 u + 2\alpha \frac{\partial u}{\partial x} + \frac{\partial^2 u}{\partial x^2}\right) + (k-1)e^{\alpha x + \beta \tau}\left(\alpha u + \frac{\partial u}{\partial x}\right) - kue^{\alpha x + \beta \tau}$$

Since, $e^{\alpha x + \beta \tau} \neq 0$

$$\beta u + \frac{\partial u}{\partial \tau} = \alpha^2 u + 2\alpha \frac{\partial u}{\partial x} + \frac{\partial^2 u}{\partial x^2} + (k-1)\left(\alpha u + \frac{\partial u}{\partial x}\right) - ku$$

Choose $\beta = \alpha^2 + (k-1)\alpha - k$ and $0 = 2\alpha + (k-1)$, so that the PDE assumes the standard parabolic form

$$\left(\beta - \alpha^2 - (k-1)\alpha + k\right)u + \frac{\partial u}{\partial \tau} = \alpha^2 u + (2\alpha + (k-1))\frac{\partial u}{\partial x} + \frac{\partial^2 u}{\partial x^2}; -\infty < x < \infty, \tau > 0$$

The argument follows from the fact that $u$ & $\frac{\partial u}{\partial x}$ should vanish and so we have,

$$\alpha = -\frac{1}{2}(k-1), \qquad \beta = -\frac{1}{4}(k+1)^2$$

Therefore,

$$\upsilon = e^{-\frac{1}{2}(k-1)x - \frac{1}{4}(k+1)^2 \tau} u(x,\tau)$$

Where

$$\frac{\partial u}{\partial \tau} = \frac{\partial^2 u}{\partial x^2}; \ -\infty < x < \infty, \tau > 0,$$

with

$$u(x,0) = u_0(x) = \max\left(e^{\frac{1}{2}(k+1)x} - e^{\frac{1}{2}(k-1)x}, 0\right) \qquad \text{(iii)}$$

This is in-fact, the pay-off function for the BS Differential equation.

Let's denote the Fourier transform of $u(x,\tau)$ by

$$u(x,\tau) = \frac{1}{2\sqrt{\pi\tau}} \int_{-\infty}^{\infty} u(x,\tau) e^{i\lambda x^2} dx$$

and assume $u_0(x)$ has a Fourier transform and $u$, $\frac{\partial u}{\partial x}$ vanish at $\infty$ so that the following results might be used

$$\frac{1}{\sqrt{2\pi}} \int_{-\infty}^{\infty} f^n(x) e^{i\lambda x} dx = (-i\lambda)^n F(\lambda); \ n = 1,2,3,\ldots\ldots \qquad \text{(iv)}$$

Multiplying equation (ii), throughout by $\frac{1}{\sqrt{2\pi}} e^{i\lambda x}$ and integrating w.r.t. $x$ from $-\infty$ to $\infty$ and using equation (iv), we obtain

$$\frac{1}{\sqrt{2\pi}} \int_{-\infty}^{\infty} u_\tau e^{i\lambda x} dx + \lambda^2 u(\lambda,\tau) + \lambda^2 u(\lambda,\tau) + \lambda^2 u(x,t) = 0 \qquad \text{(v)}$$

and equation (iii) leads to

$$u(\lambda,0) = F(\lambda) = \frac{1}{\sqrt{2\pi}} \int_{-\infty}^{\infty} e^{i\lambda x} u_0(x) dx$$

The solution to the Initial Value Problem is

$$u(\lambda,\tau) = F(\lambda) e^{-\lambda^2 \tau}$$

To find $u(x,\tau)$, apply inverse Fourier Transform and thus we obtain;

$$u(x,\tau) = \frac{1}{\sqrt{2\pi}} \int_{-\infty}^{\infty} e^{-i\lambda x - \lambda^2 \tau} F(\lambda) d\tau$$

$$= \frac{1}{\sqrt{2\pi}} \int_{-\infty}^{\infty} \int_{-\infty}^{\infty} e^{-i\lambda(x-s) - \lambda^2 \tau} u_0(s) d\tau ds \qquad \text{(vi)}$$

Next, evaluate the inner integral,

$$\int_{-\infty}^{\infty} e^{-i\lambda(x-s) - \lambda^2 \tau} d\tau = \int_{-\infty}^{0} e^{-i\lambda(x-s) - \lambda^2 \tau} d\lambda + \int_{0}^{\infty} e^{-i\lambda(x-s) - \lambda^2 \tau} d\lambda$$

$$= 2 \int_{0}^{\infty} e^{-\lambda^2 \tau} \cos[\lambda(x-s)] d\lambda$$

The last integral can be evaluated explicitly. Let the integral $I(\alpha)$ be defined as

$$I(\alpha)\big|_{\alpha = x-s} = 2 \int_{0}^{\infty} e^{-\lambda^2 \tau} \cos(\alpha \lambda) d\lambda;$$

Note that the integrand is exponentially decaying and hence bounded above.

Differentiating under the integral sign,

$$\frac{dI(\alpha)}{d\alpha} = -2 \int_{0}^{\infty} \lambda e^{-\lambda^2 \tau} \sin(\alpha \lambda) d\lambda - \frac{1}{\tau} \int_{0}^{\infty} \sin(\alpha \lambda) d\left(e^{-\lambda^2 \tau}\right)$$

$$= -\frac{\alpha}{2\tau} I(\alpha)$$

and

$$I(0) = 2 \int_{0}^{\infty} e^{-\lambda^2 \tau} d(\lambda) = \sqrt{\frac{\pi}{\tau}}$$

Combining,

$$I(\alpha)\big|_{\alpha=x-s} = 2\int_0^\infty e^{-\lambda^2 \tau} \cos(\lambda(x-s))d\lambda$$

$$= \sqrt{\frac{\pi}{\tau}} e^{-(x-s)^2/4x} \qquad \text{(vii)}$$

Using equation (vii)

$$u(x,\tau) = \frac{1}{\sqrt{4\pi\tau}} \int_{-\infty}^{\infty} u_0(s) e^{-(x-s)^2/4\tau} ds$$

$$\therefore u(x,\tau) = \frac{1}{2\sqrt{\pi\tau}} \int_{-\infty}^{\infty} u_0(s) e^{-(x-s)^2/4\tau} ds \qquad \text{(viii)}$$

where $u_0(x)$ is given by equation (iii). Again using the change of variable,

$$x' = (s-x)/\sqrt{2\tau}$$

we have,

$$u(x,\tau) = \frac{1}{2\sqrt{\pi}} \int_{-\infty}^{\infty} u_0(x'\sqrt{2\tau} + x) e^{-\frac{1}{2}x'^2} dx'$$

$$u(x,\tau) = I_1 - I_2$$

where

$$I_1 = \frac{1}{2\sqrt{\pi}} \int_{-x/\sqrt{2\tau}}^{\infty} e^{\frac{1}{2}(k+1)(x+x'\sqrt{2\tau}) - \frac{1}{2}x'^2} dx'$$

$$\therefore I_1 = e^{\frac{1}{2}(k+1)x + \frac{1}{4}(K+1)^2 \tau} N(d_1)$$

where

$$d_1 = \frac{x}{\sqrt{2\tau}} + \frac{1}{2}(k+1)\sqrt{2\tau}$$

and

$$N(d_1) = \frac{1}{\sqrt{2\pi}} \int_{-\infty}^{d_1} e^{-\frac{1}{2}s^2} ds$$

is the cumulative distribution function for the normal distribution. The calculation of $I_2$ is identical to that of $I_1$ except that $(k+1)$ is replaced by $(k-1)$ throughout.

Now, substituting from equation (i) to recover

$$C(S,t) = SN(d_1) - Ee^{-r(T-t)}N(d_2) \qquad \text{(ix)}$$

where

$$d_1 = \frac{\log(S/E) + \left(r + \frac{1}{2}\sigma^2\right)(T-t)}{\sigma\left(\sqrt{T-t}\right)}$$

$$d_2 = \frac{\log(S/E) + \left(r - \frac{1}{2}\sigma^2\right)(T-t)}{\sigma\left(\sqrt{T-t}\right)}$$

The corresponding calculation for a European Put Option follows similar lines. Its transformed pay-off is

$$u(x,0) = \max\left(e^{\frac{1}{2}(k-1)z} - e^{\frac{1}{2}(k+1)x}, 0\right) \quad \text{(x)}$$

and can be computed as above. However, having evaluated the Call, a simpler way is to use the put-call parity formula

$$C - P = S - Ee^{-r(T-t)}$$

for the value $P$ of a Put, given the value of the Call. This yields

$$P(S,T) = Ee^{-r(T-t)}N(-d_2) - SN(-d_1)$$

# 4. Software Requirement Specification:

## 4.1 Definitions/Terminologies:

Writer: The person *selling* the asset.

Holder: The person *buying* the asset.

Expiry Date: At a *prescribed time* in the future, the holder/writer of the option may purchase/sell a prescribed asset.
Underlying Asset: The holder/writer may purchase/sell a *prescribed asset* at a prescribed time.

Exercise Amount/Strike Price: The holder/writer purchases a prescribed asset at a prescribed time at a *prescribed amount.*

Arbitrage: In financial terms, there are never any opportunities to make an instantaneous risk-free profit.

Risk: Risk is commonly of two types – specific and non-specific, specific risk is the component of risk associated with a single asset, whereas non-specific risk is associated with factors affecting the whole market.

## 4.2 Notation table:

- $t$ - Time
- $S_0$ - Value of the asset at $t_0 = t$
- $S$ - Asset Price
- $E$ - Exercise Price
- $d*$ - Change in any quantity over a time interval
- $\mu$ - Measure of the average rate of growth of the asset price
- $\sigma$ - Volatility, measures the standard deviation of the returns
- $dX$ - Random variable, drawn from a normal distribution
- $\phi$ - A random variable from a standardized normal distribution
- $F(*)$ - Function
- $r$ - Risk-free interest rate
- $V$ - Value of an option, a function of $S$ and $t$
- $C$ - Call Option function
- $P$ - Put Option function

## 4.3 Hardware and Software Requirements

A simple computer with the following hardware and software features.

► Intel Pentium 4 or more.

► RAM speed of 512 MB or more.

► 1.5 GHz or faster processor.

An interactive development environment (IDE) such as CodeBlocks for building, compiling and executing the C language code.

**4.3.1 User Interfaces:** HTML 5.0 CGI platform

**4.3.2 Operating System:** Windows platform, Linux and Vista OS

**4.3.3 Memory Constraints:** Minimum memory requirements with RAM or secondary memory of about 1GB.

**4.3.4 Software System Attributes**

Reliability: This implementation is based on the Black-Scholes and Crank Nicolson's Equations; it is mathematically sound with no error included during coding. This does not slow down the system or does not cause any harm to the environment.

Availability: This software is built with a simple user interface and easily available resources.

Portability: Since this product uses platform independent HTML and C codes, it is portable.

## 5. Design:

### 5.1 Algorithm:

Finite Difference methods to solve the Black-Scholes PDE equation.

Input: Call/Put option (opt), Agreed Exercise money (E), Continuously compounded Interest rate(r), Expiry date (T).
Output: 3D shaded surface plot
step1: start
step2: read opt to find the type of option (0-call,1-put)
    read E
    read r
    read T
    for i=1 to 5 do
    begin
        for j=1 to 5 do
        begin
            a[i][j]=rand();
        end
    end
    for i=1 to 5 do
    begin
        for j=1 to 5 do
        begin
            sig[i][j]=a[i][j]
            Assign integer 11 to Nx
            Assign integer 29 to Nt

                    Assign integer 10 to L
                    Call the CrankNicholsonMethod function and pass the parameters- opt , E ,sig[i][j], r, T, Nx, Nt ,L
                    Assign the value returned to u2
                    Call the surf function by passin the parameter u2 to plot 3d shaded surface plot
            end
      end
      read xlabel, ylabel, zlabel

Step 3: Stop

## Crank Nicholson Method

Pricing a European option using the Crank-Nicolson method on the Black-Scholes PDE
Input:
Output:
Step1: start
Step 2: initialize function [U] =
      CrankNicholsonMethod (type,k,sigma2,T,N,M,Xmax)
      Assign T/N to fdeltat
      Assign Xmax/M to deltaX
      X and U are empty matrices
      Assign temp to an array of all zeroes
      for i = 1 to M+1 do
      begin
            use U to compute the pay off matrix
            if type = 0
                for j = 2 to N+1 do
                begin
                    U[1,j]=0
                    U[M+1,j] = Xmax - K*exp(-r*((j-1)*deltat))
                end
            else
                for j = 2 to N+1 do
                begin
                    U[1,j] = K
                    U[M+1,j]= 0
                end
      end
      compute the matrices and diagonalize

```
        for j = 1 to N do
        begin
                matrixB = U[:,j]
                matrixC[1,1] = temp[1,1]*U[1,j+1]
                matrixC[M-1,1] = temp[M-1,M+1]*U[M+1,j+1]
                known = matrixA*matrixB - matrixC
                Assign U_jplus1 with zeroes from [M-1,1]
                U_jplus1 = inverse of matrixD*known
                for k = 1 to M-1 do
                begin
                        U[k+1,j+1]= U_jplus1[k,1]
                end
        end
```
Assign to y the return value of the function payoff that takes type, z, K as the arguments
if type = 0
    Assign to y maximum value between z-K and 0, which symbolizes the call option
else
    Assign to y maximum value between K-z and 0, which symbolizes the put option

Step3: stop

## 5.2 Control Flow Diagram:

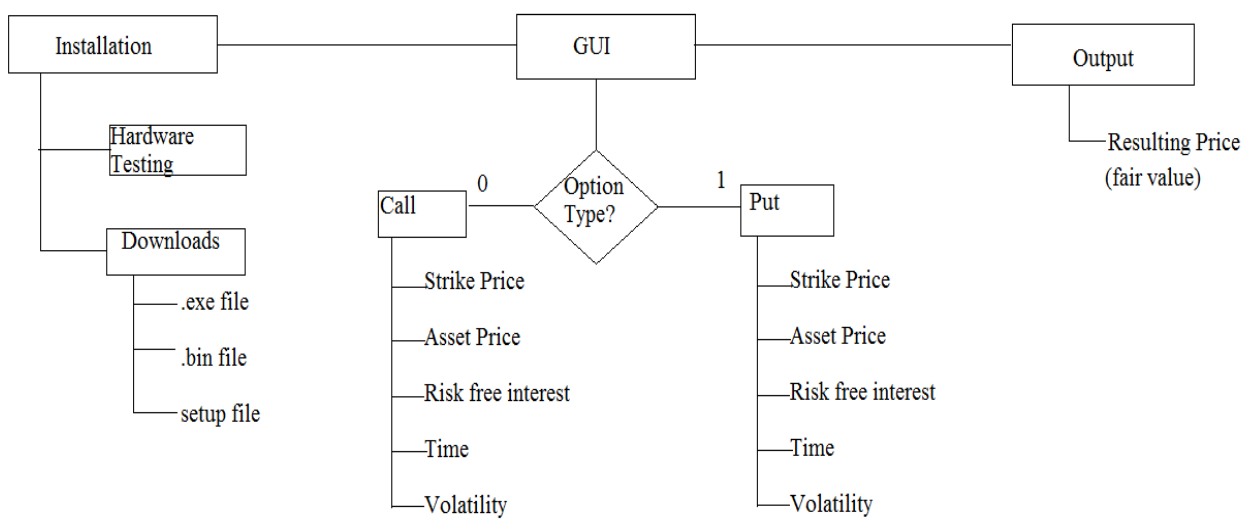

**Options Premium Calculator**

Spot Price (Rs) [    ]    Strike Price (Rs) [    ]
Interest Rate (%) [    ]  Volatility (%) [    ]
Purchase Date  6 ▼ Feb ▼ 2014    Expiry Date  6 ▼ Feb ▼ 2014
Display ● Call Option Price  ○ Put Option Price  [Results]

### 5.2.1 Call Option:

**Options Premium Calculator**

**Your Inputs**

| Strike Price (Rs) | Spot Price (Rs) | Time (Days) | Volatility (%) | Interest (%) | Type |
|---|---|---|---|---|---|
| 120 | 100 | 89 | 0.5 | 2 | Call Option |

**Results**

| Price (Rs) |
|---|
| -19.42 |

### 5.2.2 Put Option:

**Options Premium Calculator**

**Your Inputs**

| Strike Price (Rs) | Spot Price (Rs) | Time (Days) | Volatility (%) | Interest (%) | Type |
|---|---|---|---|---|---|
| 120 | 100 | 89 | 0.5 | 2 | Put Option |

**Results**

| Price (Rs) |
|---|
| 0.00 |

# 6. Bibliography:

[1] Snehanshu Saha, Swati Routh, Bidisha Goswami; Modelling Vanilla Option Prices: A Simulation Study by an Implicit Method—J. Advances in Mathematics, Vol 6, #1, pp 834-848

[2] Black, Fischer; Scholes, Myron, "The Pricing of Options and Corporate Liabilities", Journal of Political Economy 81, No. 3 (May-June 1973), pp. 637-654.

[3] Merton, R. (1973). 'Theory of Rational Option Pricing', The Bell Journal of Economics and Management Science 4, No. 1(Spring 1973), pp. 141-183.

[4] Hull, J.C. (2005), Options, Futures, and Other Derivative Securities, Prentice Hall Int. Inc., London, 1989.

[5] Black, F., (1989), How We Came Up with the Option Formula, The Journal of Portfolio Management, 15, 4-8.